\documentclass[letterpaper,twocolumn,10pt]{article}

\PassOptionsToPackage{dvipsnames}{xcolor}
\usepackage{usenix-2020-09,epsfig,endnotes}

\usepackage{authblk}
\usepackage{xspace}
\usepackage{cite}
\usepackage{amsmath}
\usepackage{bold-extra}

\usepackage{etoolbox}
\makeatletter
\preto{\@verbatim}{\topsep=1pt \partopsep=1pt}
\makeatother

\usepackage[compact]{titlesec}

\usepackage{titling}
\usepackage[ruled,linesnumbered]{algorithm2e}
\SetArgSty{textup}

\usepackage{enumitem}
\setlist{leftmargin=*,noitemsep,nolistsep}
\SetLabelAlign{center}{\strut\smash{\parbox[t]\labelwidth{\centering#1}}}

\usepackage{caption}
\usepackage[labelformat=simple]{subcaption}

\usepackage{tikz}
\usetikzlibrary{shapes.geometric}
\usetikzlibrary{arrows.meta}

\newcommand{\paratitle}[1]{\vspace{5pt}\noindent\textbf{#1}.}
\newcommand*\circlew[1]{\tikz[baseline=(char.base)]{
    \node[shape=circle,draw,inner sep=.5pt](char){\textcolor{black}{\small #1}}}}
    
\newcommand{\complexcap}[2]{\caption[#1]{\textbf{#1.} \textit{#2}}}
\newcommand{\complexcapoffig}[2]{\captionof{figure}{\textbf{#1.} \textit{#2}}}

\newcommand{\bulletcircle}[0]{\tikz[baseline=-0.5ex]{
    \node[shape=circle,draw,thick,inner sep=0pt,minimum size=10pt]{}}}
\newcommand{\bulletdiamond}[0]{\tikz[baseline=-0.5ex]{
    \node[shape=diamond,draw,thick,inner sep=0pt,minimum size=10pt]{}}}
\newcommand{\bulletgraybox}[0]{\tikz[baseline=-0.5ex]{
    \node[shape=rectangle,line width=0,inner sep=0pt,minimum size=10pt,fill=black!10!white]{
        \texttt{\string{\string}}}}}
\newcommand{\bulletarrow}[0]{\tikz[baseline=-0.6ex]{
    \draw[-Triangle,line width=0.6pt] (0,0) -- ++(10pt,0)}}
\newcommand{\bulletarrowweak}[0]{\tikz[baseline=-0.6ex]{
    \draw[-{Triangle[open]},line width=0.6pt,dotted] (0,0) -- ++(10pt,0)}}

\newcommand{\foreactor}[0]{Foreactor\xspace}
\newcommand{\iouring}[0]{\texttt{io\_uring}\xspace}

\begin{document}

\date{}

\title{\Large \bf \foreactor: Exploiting Storage I/O Parallelism with Explicit Speculation}


\author{
    Guanzhou Hu\vspace{-8pt}\\
    UW--Madison\\
    guanzhou.hu@wisc.edu
    \and
    Andrea C. Arpaci-Dusseau\vspace{-8pt}\\
    UW--Madison\\
    dusseau@cs.wisc.edu
    \and
    Remzi H. Arpaci-Dusseau\vspace{-8pt}\\
    UW--Madison\\
    remzi@cs.wisc.edu
}

\maketitle
\vspace*{-42pt}


\begin{abstract}
    We introduce \textit{explicit speculation}, a variant of I/O speculation technique where I/O system calls can be parallelized under the guidance of explicit application code knowledge. We propose a formal abstraction -- the \textit{foreaction graph} -- which describes the exact pattern of I/O system calls in an application function as well as any necessary computation associated to produce their argument values. I/O system calls can be issued ahead of time if the graph says it is safe and beneficial to do so. With explicit speculation, serial applications can exploit storage I/O parallelism without involving expensive prediction or checkpointing mechanisms.

Based on explicit speculation, we implement \foreactor, a library framework that allows application developers to concretize foreaction graphs and enable concurrent I/O with little or no modification to application source code. Experimental results show that \foreactor is able to improve the performance of both synthetic benchmarks and real applications by significant amounts (29\%-50\%).

\end{abstract}

\section{Introduction}

I/O-intensive applications, ranging from command-line file utilities to complex databases and key-value stores, play significant roles in today's computing infrastructure. They spend the majority of time in I/O system calls, accompanied with a relatively small amount of auxiliary compute to drive the system calls based on their internal data structures. For example, we observe that a \texttt{Get} operation in LevelDB~\cite{leveldb}, an LSM-tree-based key-value store, spends roughly 75\% of time in \texttt{pread} system calls on an NVMe NAND solid-state drive (SSD) when memory capacity is limited.



\begin{figure}[t]
    \centering
    \includegraphics[width=0.98\linewidth]{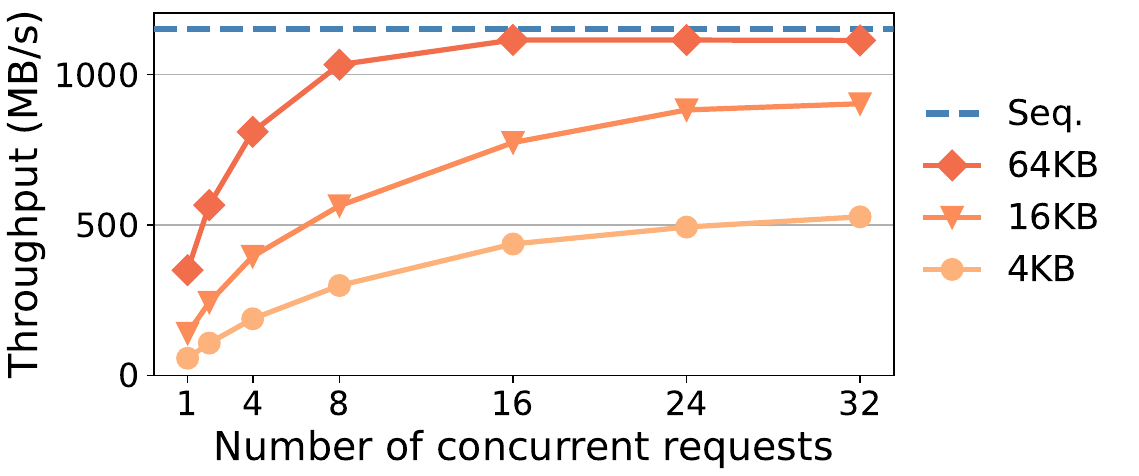}
    \complexcap{Throughput vs. I/O Concurrency on single NVMe SSD}{Dashed blue line shows sequential access steady throughput; solid lines show random mixed read-write throughput with different request sizes.}
    \label{fig:throughput-vs-io-concurrency}
    \vspace*{-8pt}
\end{figure}

On the hardware side, the evolution from legacy hard disk drives (HDDs) to modern NVMe SSDs on PCIe links has dramatically reduced device latency and improved maximum bandwidth. A hidden trend alongside the advances in storage media is that devices are offering a higher degree of internal parallelism. Modern SSDs are composed of multiple independent units, in layers of channels, packages, dies, and planes. Moreover, a single machine could be equipped with an array of multiple storage devices. Combined with an increasing number of CPU cores and larger memory capacity, it is often desired to maintain a sufficient amount of concurrent I/O requests in parallel to fully utilize storage device bandwidth. We refer to this effect as \textit{storage I/O parallelism} and visualize an example of it on an NVMe SSD in Figure~\ref{fig:throughput-vs-io-concurrency}.


I/O speculation is an ideal technique to exploit I/O parallelism by issuing I/O requests speculatively ahead of time. Current I/O speculation techniques can be categorized into speculative execution and speculative prefetching, each having its limitations. Speculative execution~\cite{spechint-kernel, speculator, xsyncfs, fast-track, prospect} moves slow I/O requests to background threads and uses predicted results to continue the application's foreground execution. It is mostly applied to write-only workloads whose results are easier to predict and requires a rollback mechanism in case of wrong prediction. Speculative prefetching~\cite{spechint, specprefetch, informed-prefetch, compiler-inserted-prefetch} guesses what parts of data are likely to be read in the near future and prefetches them in the background. It is restricted to read-only requests in a cache system context.


We propose \textit{explicit speculation}, where I/O system calls are speculatively issued under the guidance of explicit knowledge derived from application code analysis. We summarize such explicit knowledge into \textit{foreaction graphs}, a formal abstraction that describes the exact order of I/O system calls that could happen within an application function, as well as any necessary computation required to produce their argument values. An I/O system call can be issued ahead of time if the graph says it is safe and beneficial to do so. With the help of foreaction graphs, explicit speculation does not predict system call results, does not require a checkpointing mechanism, can be applied to a wider range of I/O system calls, and does not assume a cache system context.

Based on the explicit speculation technique, we implement \foreactor, a library framework that brings concurrent I/O into applications with little or no modification to their source code. To enable explicit speculation for a chosen application function, the developer draws its foreaction graph, describes it as a plugin code file to be put alongside application source code, compiles it with the application, and pre-loads the libforeactor dynamic library when running the application. \foreactor intercepts POSIX library invocations on system calls and consults the foreaction graph for submitting the next few system calls that are foreseen to happen. \foreactor incorporates both \iouring, a recent Linux kernel asynchronous I/O interface, and a general user-level thread pool as the backend engine to efficiently delegate I/O system calls to background threads.


We evaluate \foreactor with real application workloads and show that it improves the performance of the \texttt{du} command-line utility by 50\%, the \texttt{cp} utility by 29\%, B+-tree scan and bulk-loading by up to 37\%, and the LevelDB Get operation with YCSB by up to 34\%. We also demonstrate the benefits of separating the foreaction graph abstraction from specific backend implementations; with a seamless switch to a user-level thread pool backend, Foreactor achieves an 18\% speedup for B+-tree bulk workloads.



The rest of this paper is organized as follows. \S\ref{sec:background} provides background on storage I/O parallelism and existing software techniques to exploit this effect. \S\ref{sec:technique} describes explicit speculation and defines the foreaction graph abstraction. \S\ref{sec:app-study} presents a case study of a few suitable application workloads. \S\ref{sec:design-impl} describes the design and implementation of the \foreactor framework. \S\ref{sec:evaluation} presents experimental evaluation results of \foreactor. \S\ref{sec:discussion} provides additional discussion on some practical aspects. \S\ref{sec:related-work} summarizes related work, and \S\ref{sec:conclusion} concludes the paper.

\section{Background}
\label{sec:background}

We provide background on storage I/O parallelism and existing software techniques to exploit such parallelism, and discuss their limitations in this section.

\subsection{Storage Device Parallelism}

Persistent storage devices are often built upon multiple smaller units of storage media. In flash-based SSDs -- which have become increasingly popular in today's storage systems -- there are multiple levels of smaller units, namely \textit{channels}, \textit{packages}, \textit{dies}, and \textit{planes}, contributing to the aggregate bandwidth. The flash controller inside an SSD device can manage up to 10 independent channels. Each channel connects multiple flash memory chips, called packages. Each package is in turn composed of multiple dies (for interleaving) or planes (for true parallelization)~\cite{ssd-design, unwritten-contract-ssd}.

The maximum effective throughput out of a storage device is reached when all of its independent units are operating on data being requested by the user. To make this goal easier to achieve, data is often divided into small chunks, e.g. 512B sectors or 4KB pages, and striped across all units in such granularity. If an I/O request is large enough to span an entire stripe, then all units work together at the same time on fetching or manipulating one chunk of data, delivering full device bandwidth. However, I/O request sizes vary and are often much smaller than the size of a stripe. In general, it is required to maintain a sufficient amount of concurrent I/O requests on the fly, such that their addresses statistically cover all units of storage media. The number of concurrent I/O requests is often referred to as the queue depth (QD). Depending on the average request size of a workload, a QD of 16 or higher may be necessary to fully utilize device bandwidth~\cite{unwritten-contract-ssd, unwritten-contract-optane}.

Additional CPU power and memory capacity is required to maintain concurrent I/O requests. Luckily, modern machines are equipped with exponentially more CPU cores and larger main memory, allowing them to support higher I/O concurrency to fully exploit the effect of storage I/O parallelism as shown in Figure~\ref{fig:throughput-vs-io-concurrency}. In fact, storage systems demanding extremely high throughput often manage a RAID array of independent devices~\cite{raid, raid-ssd}, introducing yet another factor of parallelism. For simplicity, we restrict the scope of this paper to a single SSD device.

\subsection{I/O Speculation}
\label{sec:io-speculation}

I/O speculation refers to the set of I/O optimization techniques that involve speculative work, i.e., I/O requests and/or computation that are predicted to happen and are started ahead of time, but may not eventually happen in the actual execution of the application. Current I/O speculation techniques can be categorized into two classes: speculative execution and speculative prefetching.

\vspace{-2.2pt}
\paratitle{Speculative Execution} Upon submitting a slow I/O request during the foreground execution of an application, instead of completing the I/O request in-place, the system offloads it to a background thread and predicts its result, then continues foreground execution assuming the prediction is correct. Speculation on the foreground thread must be paused when it reaches the point of returning an externally visible result to the user. This technique shares much similarity with speculative execution in CPU branch prediction. The foreground thread is often named the fast-path thread, since it skips slow I/O requests and speculatively executes subsequent computation (as well as submission of later dependent I/O requests). Example realizations of this technique include SpecHint -- a framework for run-time I/O hint generation~\cite{spechint, spechint-kernel}, Speculator -- a speculative distributed file system~\cite{speculator}, Speck -- a parallel security check software~\cite{speck}, and XsyncFS -- a locally asynchronous file system batching write commits~\cite{xsyncfs}.


Speculative execution has two major limitations, preventing it from being useful for exploiting storage I/O parallelism. First, it must be able to generate a predicted result of any offloaded I/O request, which is hard for most read requests. Hence, it is often applied to compute-heavy workloads where I/O requests are rare, or to write-heavy workloads where all writes are predicted to return success. Second, it requires a checkpoint-rollback mechanism, e.g. undo logging, on all data touched by speculation, so that data and internal application states stay correct in cases of wrong predictions. Such mechanisms often introduce significant performance overhead.

\vspace{-2.2pt}
\paratitle{Speculative Prefetching} Within a cache framework context, the system makes guesses on what parts of data are going to be read in the near future and prefetches them into cache ahead of time. Read requests are either batched together or made parallel to increase overall request scale and improve bandwidth utilization. Cache read-ahead is the simplest form of speculative prefetching, where the system always favors consecutive addresses following recently accessed data. More sophisticated speculative prefetching policies, such as informed prefetching~\cite{informed-prefetch}, MPI-IO caching~\cite{specprefetch}, signature-based prefetching~\cite{sig-based-prefetch}, and speculative RPCs~\cite{specrpc} have been proposed to optimize for specific applications and workload scenarios.

Speculative prefetching naturally exploits I/O parallelism, yet has three restrictions limiting its generality. First, it can only be applied to stateless read requests, because argument values of speculatively issued requests come from guesses and may end up wrong. Any requests with side effects, e.g. writes, cannot be safely submitted ahead of time. Second, it assumes a cache system context and requires fine-tuned application-specific cache management policies to yield good performance. Deriving such policies is challenging and there is no standard methodology as guidance. Third, it often relies on run-time trace collection to produce reasonable guesses on future I/O requests, introducing considerable overhead.

\begin{figure*}[t]
    \centering
    \begin{subfigure}[t]{0.315\textwidth}
        \includegraphics[height=2.7cm]{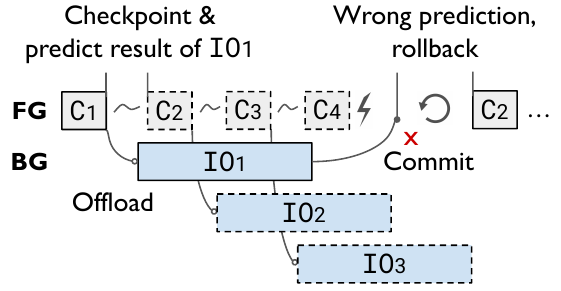}
        \captionsetup{justification=centering}
        \caption{Speculative Execution}
        \label{fig:technique-speculative-execution}
    \end{subfigure} \hfill
    \begin{subfigure}[t]{0.36\textwidth}
        \includegraphics[height=2.7cm]{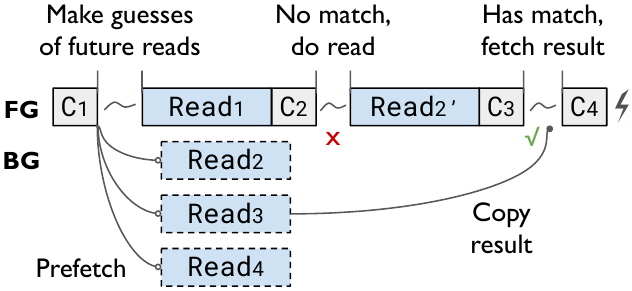}
        \captionsetup{justification=centering}
        \caption{Speculative Prefetching}
        \label{fig:technique-speculative-prefetching}
    \end{subfigure} \hfill
    \begin{subfigure}[t]{0.315\textwidth}
        \includegraphics[height=2.7cm]{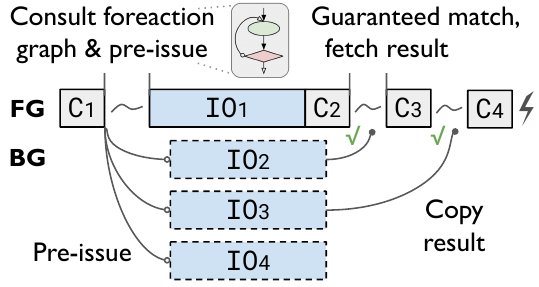}
        \captionsetup{justification=centering}
        \caption{Explicit Speculation}
        \label{fig:technique-explicit-speculation}
    \end{subfigure}%
    \complexcap{Demonstration of I/O Speculation Techniques}{Gray boxes represent computation. Blue boxes represent I/O requests. Dashed border means speculative content. Lightning symbol represents the return point of application function. FG stands for foreground application thread. BG stands for background I/O threads.}
    \label{fig:techniques}
\end{figure*}

\subsection{Asynchronous I/O and \textbf{\iouring}}

Besides I/O speculation techniques, applications may also be re-programmed to perform asynchronous I/O explicitly. This can be done either though manually implementing a user-level thread pool I/O engine, or by using an asynchronous I/O interface provided by the underlying operating system. A system supporting asynchronous I/O often exposes a separate pair of interfaces for I/O submission and completion, respectively. For example, the legacy Linux \texttt{aio} interface uses \texttt{io\_submit()} for submissions and \texttt{io\_getevents()} for blocking and waiting on completions. However, \texttt{aio} does not support decent I/O parallelization in the background.


Linux kernel version 5.1 introduces \iouring~\cite{io-uring}, a modern alternative that supports more I/O system calls, has no restrictions on request alignment and page cache configuration, and enables efficient concurrent I/O with the help of an in-kernel thread pool named \texttt{io\_}workqueue. \iouring exposes a dual ring-queue interface, one as the submission queue (SQ) and one as the completion queue (CQ). Application calls \texttt{io\_uring\_setup()} system call once to allocate space for the two ring-queues. To submit a batch of I/O requests, the application fills a corresponding number of entries after the current SQ tail pointer, and makes a single \texttt{io\_uring\_enter()} system call to inform the kernel about the new entries. Submitted requests may be carried out either serially by one kernel thread or in parallel by multiple \texttt{io\_}workqueue threads. An extra \textit{linking} feature is provided to link requests into a chain to force them to be executed in order. To check for completions, no system calls are required and the application simply checks the position of the CQ tail pointer, which is updated by the kernel whenever a new completion entry gets filled.

A primitive asynchronous I/O interface alone is not sufficient for applications to benefit from I/O asynchrony and I/O parallelism. Developers still have to do significant re-engineering of application source code to decide when and where to submit I/O requests and wait for completions. This process is largely application-specific and platform-specific -- such code refactoring has to be repeated for every new application and on every different asynchronous I/O backend.

\section{Explicit Speculation}
\label{sec:technique}

We propose \textit{explicit speculation}, a form of I/O speculation guided by explicit knowledge derived from application code analysis. We formalize the result of code analysis as a new abstraction -- the \textit{foreaction graph} -- which is a directed graph describing the order of all I/O-related system calls that a chosen application function could possibly issue, as well as the sources of their argument values. With the help of the foreaction graph abstraction, explicit speculation enables a general method to decide on candidates of speculative I/O requests, to pre-compute necessary argument values at appropriate times, and to maintain a sufficient amount of I/O concurrency to fully exploit storage I/O parallelism.


\subsection{Motivation and Goals}

The idea of explicit speculation is motivated by the following four facts. First, I/O parallelism is currently under-exploited. Most I/O acceleration techniques, e.g. kernel-bypassing, focus on reducing software stack latency. Existing I/O speculation techniques improve I/O parallelism in workload-specific ways and introduce considerable overhead. Primitive asynchronous I/O interfaces require significant development effort re-writing applications to make use of them.

Second, most application code logic is naturally written in a serial way and contains an explicit pattern of I/O system calls. It should be possible to formalize such knowledge accurately for general application functions and use this explicit knowledge to enable I/O speculation with low overhead.

Third, the order of I/O requests, their argument values, and their mutual dependencies are the most critical information needed to enable efficient I/O speculation. However, no systematic methodology has been proposed to guide concurrent I/O based on these information.

Fourth, the appearance of Linux \iouring now allows us to batch and parallelize I/O system calls in an asynchronous manner flexibly and efficiently.


Accordingly, by proposing explicit speculation, we aim to achieve these goals:
\begin{itemize}
    \item Define a formal abstraction, named the foreaction graph, that statically describes all the necessary information needed of an application function to make I/O speculation decisions. (\S\ref{sec:foreaction-graph} and \S\ref{sec:principles-of-speculation})
    \item Conduct case studies of code analysis on a set of applications that are suitable for concurrent I/O and derive their foreaction graphs. (\S\ref{sec:app-study})
    \item Design and implement a general library framework that brings concurrent I/O into serially-written applications with the knowledge of foreaction graphs to exploit storage I/O parallelism, while introducing minimal modification to their source code. (\S\ref{sec:design-impl})
\end{itemize}

\subsection{Foreaction Graph}
\label{sec:foreaction-graph}

We now define the foreaction graph abstraction formally. We make two implicit design decisions. First, we consider system calls as the granularity of I/O requests, because they naturally capture application code logic and sit on the clear boundary between userspace and the kernel. Second, we consider individual application functions, e.g. the \texttt{DBImpl::Get()} code path of LevelDB, as the scope of speculation. Full application analysis introduces too much complexity and is impractical for modern I/O-intensive applications. Also, their bottlenecks typically occur in just a few critical-path functions that involve bulk operations.

To enable issuing I/O system calls ahead of time, we found that the following information about a chosen function's code logic are required: \circlew{1} the collection of possible I/O system calls and their original order, \circlew{2} the argument values of each system call instance, and \circlew{3} the dependencies between system calls, if any. We organize these information in the form of a directed graph with annotations. Figure~\ref{fig:foreaction-graph-example} shows the foreaction graph of an example application function that reads the first block of each file in a list of input files and may return early when e.g. a search pattern is found.


\paratitle{Foreaction Graph Structure} Specifically, a foreaction graph is composed of the following elements:
\begin{itemize}[leftmargin=20pt,align=center,wide=0pt,itemsep=5pt,topsep=5pt]
    \item[\bulletcircle] \textit{Syscall nodes} represent I/O system calls instances and each have a certain type, e.g. \texttt{pread} or \texttt{pwrite}. A syscall node typically maps to a unique system call invocation in application source code. A syscall node is \textit{pure} if it is a read-only system call that has no side effect other than possibly bringing data into the OS page cache, such as \texttt{pread and \texttt{fstat}}. We translate any \texttt{read} with implicit offset into a \texttt{tell}-\texttt{pread}-\texttt{seek} combination and consider it pure as well. Non-pure system calls, e.g. \texttt{pwrite}, will leave permanent side effects and thus cannot be arbitrarily issued ahead of time.
    \item[\bulletdiamond] \textit{Branching nodes} represent split points into multiple possible execution paths containing different syscall node sequences. A branching node typically maps to a branching point in application code control flow, such as an \texttt{if} condition or a loop boundary. Note that only branching points that lead to different I/O system call sequences appear as branching nodes in the graph. An \texttt{if-else} block of pure computation, for example, will not appear in the graph structure.
    
\begin{figure}[!t]
    \centering
    \includegraphics[width=\linewidth]{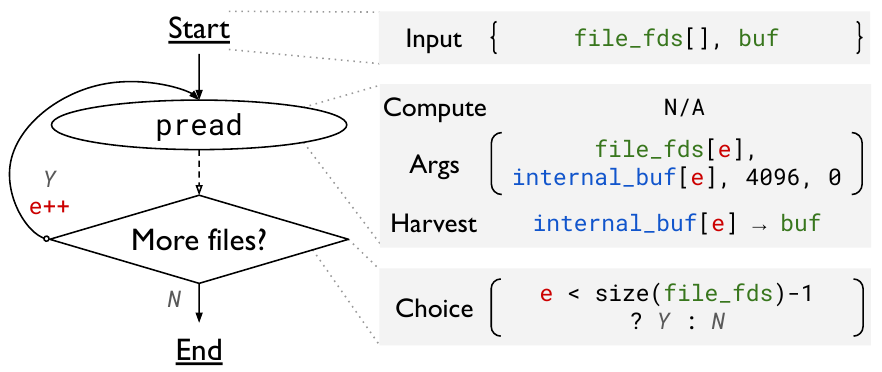}
    \complexcap{Foreaction Graph Example}{}
    \label{fig:foreaction-graph-example}
    \vspace*{-10pt}
\end{figure}
    
    \item[\underline{S}] \textit{Start node} is a special node representing the entrance of the function and appears exactly once in each graph.
    \item[\underline{E}] \textit{End node} is a special node representing the exit of the function and appears exactly once in each graph.
    \item[\bulletarrow] \textit{Edges} connect nodes together to form a directed graph. Let us first assume a graph without loops. In this case, the graph must be a directed acyclic graph (DAG) flowing from the start node as source to the end node as drain. The start node has no incoming edge and exactly 1 outgoing edge; each syscall node has 1 or more incoming edges and exactly 1 outgoing edge; each branching node has 1 or more incoming edges and 1 or more outgoing edges; the end node has 1 or more incoming edges and no outgoing edge. Based on this setup, we introduce \textit{loops} into the graph by allowing an outgoing edge from a branching node to be a looping-back edge pointing to a prior syscall node or a prior branching node. Every such looping-back edge is associated with an \textit{epoch number}, explained later in this section, that records the number of traversals across it.
    \item[\bulletarrowweak] An edge could be marked as \textit{weak} (and drawn as a dashed arrow), meaning the function could possibly exit early at some point during its execution on that edge. This denotation is syntactically equivalent to having an extra branching node on the edge with two outgoing edges, one to its original successor node and one to the end node.
    \item[\bulletgraybox] \textit{Annotations} are critical information attached to nodes that specify all the necessary computation required to support I/O speculation on the graph. We explain annotations in more detail below.
\end{itemize}

\paratitle{Node Annotations} Annotations control the computation of syscall node argument values and the execution flow of the graph in case of branching. They are a crucial part of the foreaction graph abstraction because they encode the explicit knowledge we need to compute system call arguments ahead of time. Each node type has its own set of annotations.

A \textit{start node} has one section in its annotation, \textit{Input}, which is a collection of application variables whose values may be used by other nodes in the graph.

\textit{Syscall nodes} have three sections in their annotations: \circlew{1} \textit{Compute}, which describes any necessary computation needed to produce its argument values from input application variables. \circlew{2} \textit{Args}, which is the list of arguments, according to the type of the system call, to be used when this system call instance is to be issued speculatively. \circlew{3} \textit{Harvest}, which specifies the action to be done after the completion of the system call to reflect its result. For brevity, when presenting a foreaction graph, if all the computation needed by a syscall node is simple logic such as array lookup, the Compute section may be omitted and inlined in the Args section. Compute may also contain a special operation called \textit{Link}, which forces the backend to always submit this system call together with the next one down the graph and execute them in sequential order. This is useful when, e.g., a write follows a read in a copy loop.

\textit{Branching nodes} have one section in their annotations, \textit{Choice}, which describes the computation of the branching decision, i.e., which outgoing edge to take.

    

\paratitle{Annotation Variables} The following types of variables may appear in annotations:
\begin{itemize}[leftmargin=18pt,align=center,wide=0pt,itemsep=5pt,topsep=5pt]
    \item[\texttt{\color{ForestGreen} {\large A}}] \textit{Application state} variables are ones that exist in original application code and whose values can be used directly. All application state variables used in the graph should appear in the \textit{Input} annotation of the start node.
    \item[\texttt{\color{NavyBlue} {\large I}}] \textit{Internal extra state} variables are ones added by the graph to help with the computation of argument values or to represent extra memory buffers needed by read-like system calls.
    \item[\texttt{\color{BrickRed} {\large e}}] \textit{Epoch number}s keep track of the number of loop iterations and are used to index array-like variables.
\end{itemize}

Throughout this paper, we assume that foreaction graphs are obtained from manual analysis of application code by developers. We discuss the feasibility of automated generation of foreaction graphs in \S\ref{sec:discussion}.

\begin{figure*}[t]
    \centering
    \begin{subfigure}{0.49\textwidth}
        \centering
        \includegraphics[width=0.9\columnwidth]{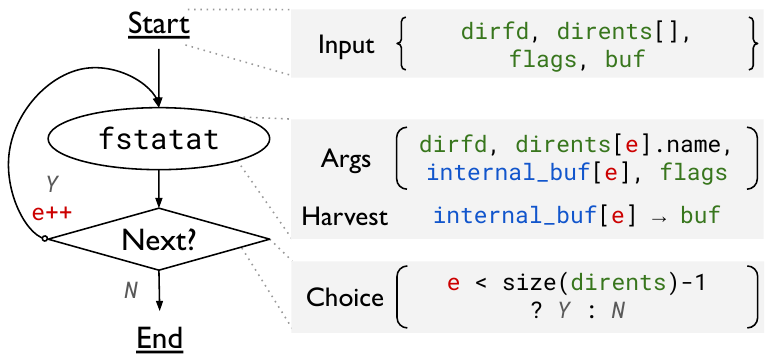}
        \captionsetup{justification=centering}
        \caption{Metadata Stat Loop}
        \label{fig:foreaction-graph-du}
        \vspace{12pt}
        \includegraphics[width=0.96\columnwidth]{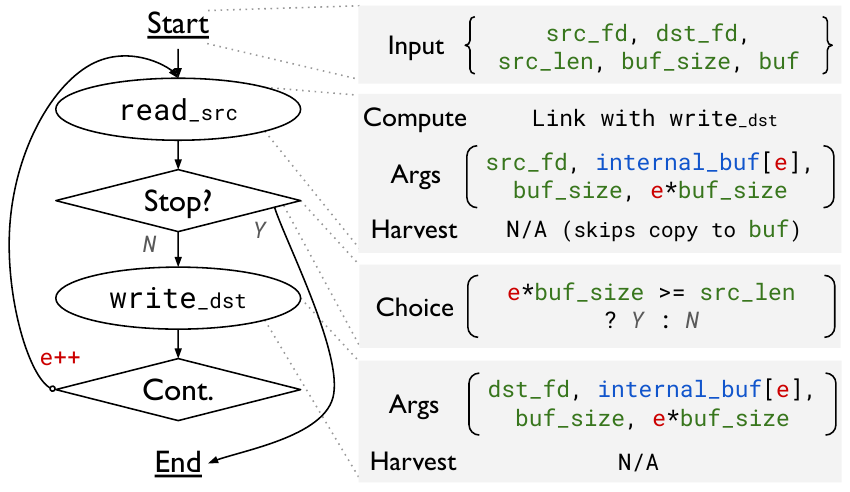}
        \captionsetup{justification=centering}
        \caption{Data Copy Loop}
        \label{fig:foreaction-graph-cp}
    \end{subfigure}%
    \begin{subfigure}{0.51\textwidth}
        \centering
        \includegraphics[width=\columnwidth]{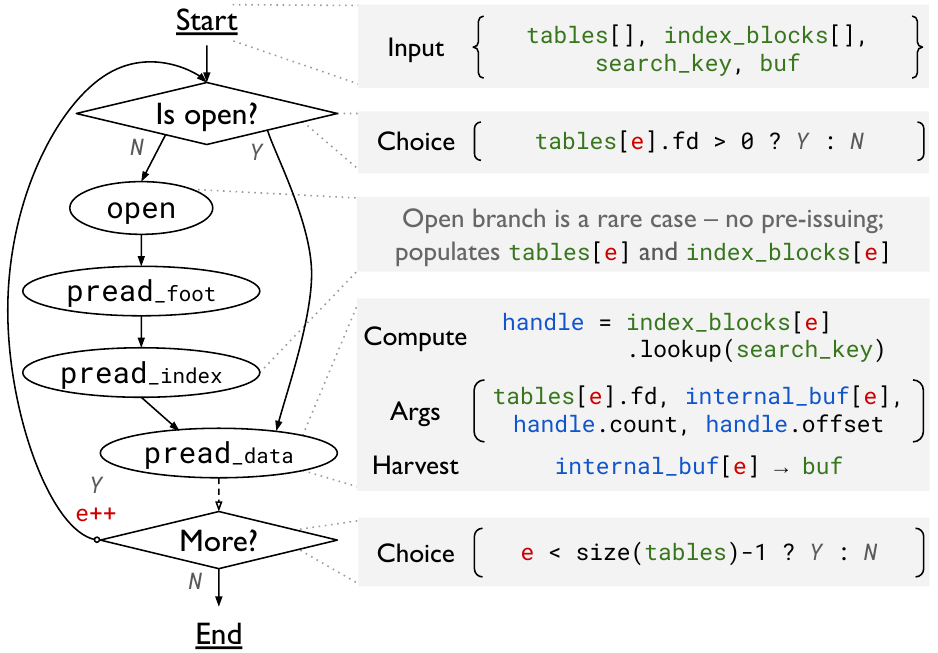}
        \captionsetup{justification=centering,aboveskip=5pt}
        \caption{LSM-tree Search}
        \label{fig:foreaction-graph-ldb}
        \vspace{10pt}
        \includegraphics[width=0.85\columnwidth]{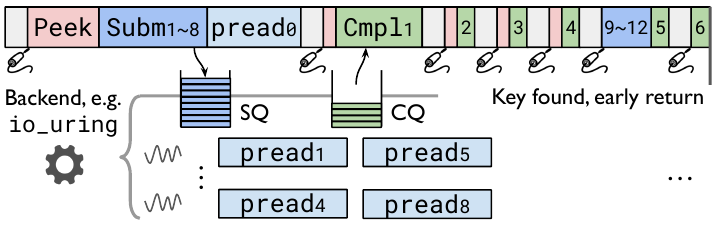}
        \captionsetup{justification=centering}
        \caption{LSM-tree Search Timeline w/ Speculation}
        \label{fig:foreaction-graph-ldb-timeline}
    \end{subfigure}%
    \complexcap{Representative Application Foreaction Graphs}{\subref{fig:foreaction-graph-du}, \subref{fig:foreaction-graph-cp}, and \subref{fig:foreaction-graph-ldb} shows the foreaction graphs for three common application workloads. \subref{fig:foreaction-graph-ldb-timeline} illustrates a possible execution timeline of \subref{fig:foreaction-graph-ldb} with explicit speculation.}
    \label{fig:foreaction-graph-apps}
\end{figure*}

\subsection{Principles of Speculation}
\label{sec:principles-of-speculation}

We then describe the principles of doing explicit speculation on a given foreaction graph and discuss the reasons behind these principles. The ultimate goal is to try to maintain a sufficient number of concurrent I/O system calls such that the aggregate request scale is large enough to yield close-to-optimal throughput, while not violating application semantic given by the foreaction graph. To achieve this goal, we obey the following guidelines.

\paratitle{Speculate at run-time} Speculation of I/O system calls should happen progressively at run-time to amortize the overhead. This means both the decision making on which systems calls to pre-issue and the submission of these system calls are carried out along the way of original function execution.

\paratitle{Peek in execution order} The best candidates for speculatively-issued system calls are those that immediately follow the current progress of execution, because their results are the soonest to be needed. Therefore, when preparing system calls for speculative submission, the best strategy is to peek at the next few successor nodes by following outgoing edges up to a certain depth.

\paratitle{Compute argument values explicitly} Having the foreaction graph abstraction gives us the advantage of knowing how to compute system call argument values explicitly, instead of relying on run-time guesses as in speculative prefetching. Therefore, compute these values explicitly whenever possible.

\paratitle{No unrecoverable side effects} Any non-pure system calls (as defined in \S\ref{sec:foreaction-graph}) may leave unrecoverable side effects when issued speculatively, such as unintended writes that do not eventually happen. Instead of implementing a checkpoint-rollback mechanism with high overhead as in speculative execution systems, we choose to put a rule that any non-pure system calls could be pre-issued only if they are guaranteed to happen. This means there are no \textit{weak} edges along the path from the current node of execution to the prepared candidate. This rule limits our ability to parallelize certain write-heavy functions, but greatly simplifies the correctness model.

\paratitle{Control depth according to scale} The depth of peeking controls the approximate number of concurrent I/O system calls. It should be set to a value such that the aggregate request scale, i.e. average request size times depth, is large enough to fully exploit device bandwidth. For example, given an SSD of stripe size 128KB and a function containing a loop of 4KB reads, an ideal depth of peeking should be larger than 32. The optimal depth varies across different hardware platforms and workloads, and may have an obvious upper bound for functions that at most contain a certain number of I/O calls.

\section{Application Case Study}
\label{sec:app-study}

We do a brief case study of several I/O-intensive application scenarios, including regular I/O loops in command-line utilities, B+-tree scan and bulk-loading, and LSM-tree search. We present their foreaction graphs to demonstrate the determinism in their system call patterns and the potential of I/O parallelism exploitation.

\subsection{Regular I/O Loops}

Functions that contain a loop of I/O requests are obvious candidates for concurrent I/O. Let us first consider a function that loops through all entries of a directory and calls \texttt{fstatat} on each entry to collect their metadata information. Figure~\ref{fig:foreaction-graph-du} draws its foreaction graph. Since all \texttt{fstatat} calls are pure and are independent with each other, they can be safely pre-issued in parallel. To do so, a set of internal buffers have to be created to hold the results from concurrent requests.

A slightly more complex case would be a loop that copies blocks of data by reading from a source file and writing into a destination file. Figure~\ref{fig:foreaction-graph-cp} draws the foreaction graph for this case. In each iteration of the loop, the \texttt{write} call depends on the \texttt{read} call to fetch data first, thus we cannot freely pre-issue the \texttt{write}s even though their argument values can be explicitly computed. The solution is to leverage the \textit{Link} feature so that each read-write pair are submitted together and are forced to execute in order. Also, note that the \texttt{read} node does nothing in its \textit{Harvest} section to avoid unnecessary memory copies -- each \texttt{write} directly uses the corresponding internal buffer that the linked \texttt{read} populates.

\subsection{B+-tree Scan and Bulk-loading}
\label{sec:app-study-bplus-tree}

The B+-tree is a well-known data structure for efficiently indexing both relational and key-value data. A B+-tree is an $m$-ary balanced search tree, where $m$ is typically at the magnitude of 100s. Every node is a fixed-sized page containing a sorted array of keys -- alongside the keys, internal nodes store page IDs of child nodes, while leaf nodes store values (or pointers to values if they reside in a separate location). Each node also maintains the page ID of its right sibling to support bulk operations. All pages of a B+-tree typically reside in a single database file on backend storage. Although an individual search or insertion is a strict pointer-chasing procedure, a scan or bulk-loading operation involves a long sequence of leaf page I/O, revealing an opportunity for I/O parallelization.

In a range scan, it first traverses the tree twice for the start key and the end key, respectively, to locate their leaf pages. A sequential scan of keys from start to end is then carried out to gather key-value pairs within the range, triggering a leaf page read upon crossing any sibling pointer. This operation can be parallelized by first looking up the last level internal pages and gathering all candidate leaf page IDs, then issuing a certain number of leaf page reads in parallel. Bulk-loading could happen upon initializing a new B+-tree from a sorted stream of incoming data. Instead of traversing the tree for every key-value pair, it simply writes them into the right-most leaf node (possibly triggering splits when a node becomes full). This leads to a loop of leaf page writes.

The leaf page I/O loop in scan and bulk-loading turns out to be straight-forward and can be modeled as a foreaction graph similar to the stat loop in Figure~\ref{fig:foreaction-graph-du}, replacing \texttt{fstatat} with \texttt{pread} or \texttt{pwrite}. Also, note that these two bulk operations typically happen upon initialization or on cold data, leading to inevitable cache misses.

\subsection{LSM-tree Search}
\label{sec:app-study-lsm-tree}

The Log-Structured Merge Tree (LSM-tree) is a widely-used data structure for storing key-value data and is optimized towards efficient inserts/updates. Searching for a key in an LSM-tree, however, may involve a long chain of \texttt{pread}s through multiple files. Consider a typical LSM-tree implementation where records are stored in sorted table files. Collections of tables form levels starting with level 0. New records (including updates to existing keys) are inserted into an in-memory table first, and when it reaches certain size limit, it is dumped as a new table file and is pushed into level 0. Tables in level 0 thus may have overlapping key ranges. When a level becomes too crowded, a background compaction procedure is scheduled to merge all records in the level and dump them as table files into the lower level. In all levels starting at 1, tables within a level cover non-overlapping key ranges.

A search operation involves first looking up the key in memory; if not found, it then needs to traverse through a list of table files. The list consists of all level-0 tables, from the newest to the oldest, with range covering the key, followed by at most one file per lower level. If a match is found along the way, the traversal terminates early. In the worst case, a search operation may involve 12$\sim$19 \texttt{pread}s in LevelDB. Each table file is roughly 2MB in size and contains a sequence of sorted records, an index block mapping keys to offsets, and a signature footer. Looking up a key in a table involves a simple computation of index block lookup followed by a \texttt{pread} into the table file for the candidate data block.

Figure~\ref{fig:foreaction-graph-ldb} draws the foreaction graph for the LSM-tree search operation. We would like to point out three things about the graph. First, the open branch is a rare case and is therefore omitted. Second, the \textit{Compute} annotation of \texttt{pread\_data} node captures the index block lookup, which in LevelDB's case is a binary search. Third, the edge from \texttt{pread\_data} to the branching node is a \textit{weak} edge, because the function may terminate early if a match is found. If explicit speculation is applied to LSM-tree search, Figure~\ref{fig:foreaction-graph-ldb-timeline} illustrates a possible execution timeline when there are 12 candidate tables and the key is found in the 7th table. The needles represent timepoints where the application calls \texttt{pread}s. Reasoning behind the timeline will be covered in \S\ref{sec:design-impl}.

\section{Foreactor: Design and Implementation}
\label{sec:design-impl}

We implement \foreactor, a userspace framework that enables explicit speculation in C/C++ applications. In this section, we describe the components of the framework, the essential pre-issuing algorithm, its correctness guarantees, and additional implementation details.

\subsection{Framework Components}
\label{sec:framework-components}

\begin{figure}[t]
    \centering
    \includegraphics[width=0.7\linewidth]{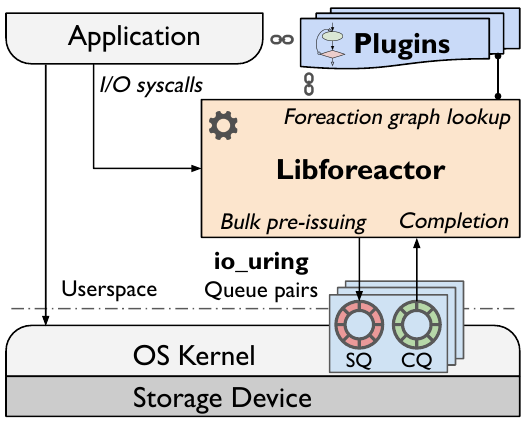}
    \complexcap{Overview of Foreactor Architecture}{}
    \label{fig:foreactor-arch}
    \vspace{-10pt}
\end{figure}

The architecture of \foreactor is shown in Figure~\ref{fig:foreactor-arch}. It is mainly composed of two parts: \circlew{1} \textit{libforeactor}, the core dynamic library that intercepts I/O system calls and controls speculation on foreaction graphs, and \circlew{2} application-specific foreaction graphs written in the form of auxiliary plugin code.

\paratitle{Libforeactor} Libforeactor is the core dynamic library that any application using \foreactor links against. The library plays four roles in enabling explicit speculation. \circlew{1} It provides an interface to plugin code for building foreaction graphs on chosen application functions and registering them for speculation. \circlew{2} It intercepts registered function calls as well as I/O system calls (POSIX library calls to be precise) to track the progress of code execution. \circlew{3} When a registered function is active, it consults the corresponding foreaction graph and pre-issues a certain number of I/O system calls ahead of time, using the algorithm explained in \S\ref{sec:pre-issuing-algo}. \circlew{4} It manages an asynchronous system call engine, by default Linux \iouring, to submit concurrent I/O system calls and harvest their completions.


\paratitle{Foreaction Graph as Plugin Code} To materialize a conceptual foreaction graph into code that libforeactor understands, a plugin file should be created and put alongside application source code. Suppose the target function to be optimized is \texttt{f}, its plugin file should contain the following content.

First, a builder function is required to programmatically compose the foreaction graph structure. Libforeactor provides APIs including \texttt{AddSyscallNode()}, \texttt{AddBranchingNode()}, \texttt{SyscallSetNext()}, as well as \texttt{BranchAppendChild()}. The builder function is invoked only once upon the first invocation of \texttt{f}.

Second, a wrapper function is needed to shadow \texttt{f} and capture the values of all necessary input variables that appear in the \textit{Input} annotation.

Third, each syscall node requires two custom helper stubs to realize their attached annotations:

{\tt \small
\begin{verbatim}
 ComputeArgs(epoch, *args...) -> (ready, link);
 SaveResult(epoch, rc) -> void;
\end{verbatim}
}

\noindent Specifically, \texttt{ComputeArgs()} implements \textit{Compute} and \textit{Args} sections of the annotation, while \texttt{SaveResult()} implements \textit{Harvest}. \texttt{ComputeArgs()} either produces the system call's argument values from current input variables, or returns non-ready to indicate that arguments values cannot be produced yet at this time point. The link flag is returned true to indicate that this system call should be linked with the next one down the graph upon submission, as explained in \S\ref{sec:foreaction-graph}. \texttt{SaveResult()} is evaluated exactly once per syscall node (and per epoch if there are loops) to retrieve its return code in case it was issued speculatively in the background.

Fourth, each branching node needs a \texttt{ComputeArgs()} stub that produces the branch decision, which corresponds to the \textit{Choice} annotation. It may also return non-ready to indicate that the branch decision cannot be made yet at this time point.

An application could have multiple functions registered with \foreactor, in which case there will be multiple foreaction graphs and hence multiple plugin files. Every foreaction graph instance is per-thread local, meaning that a multi-threaded application could have multiple threads executing the same function, each performing explicit speculation with \foreactor independently.

\paratitle{Build System Modification} To activate \foreactor, no modification to application source code is required. However, minor modification to its build system is needed to include the compilation of plugin file and the linking of it with the rest of the application. Alternatively, the application could include its plugins explicitly around wrapped functions.

\subsection{Pre-issuing Algorithm}
\label{sec:pre-issuing-algo}

\begin{algorithm}[t]
    \caption{Pre-issuing Algorithm}
    \label{alg:pre-issuing-algo}
    
    \small
    
    \KwData{peek \texttt{depth}, intercepted node \texttt{frontier}}
    
    \texttt{n = frontier.next}, \texttt{e = frontier.out\_edge}\;
    \texttt{weak = false}\;
    
    \While{\texttt{depth-- > 0} and \texttt{n} is not null}{
        if \texttt{e} is weak, set \texttt{weak = true}\; \label{alg:line:track-weak}
        
        \While{\texttt{n} is branching node}{ \label{alg:line:outer-while}
            \texttt{n} = \texttt{n.ChooseBranch()}\;
            if \texttt{n} is null, \texttt{goto} line {\hypersetup{hidelinks}\ref{alg:line:batch-submit}}\;
        }
        
        \texttt{ready = n.ComputeArgs()}\; \label{alg:line:compute-args}
        \If{\texttt{ready} and not (\texttt{weak} and \texttt{n} non-pure)}{ \label{alg:line:foreactable-cond}
            \texttt{io\_uring.Prepare(n)}\;
        }
        
        \texttt{n = n.next}, \texttt{e = n.out\_edge}\; \label{alg:line:peek-next}
    }
    \texttt{io\_uring.SubmitAllPrepared()}\; \label{alg:line:batch-submit}
    
    \uIf{\texttt{frontier} not in \texttt{io\_uring}}{
        invoke \texttt{frontier} synchronously\;
    }\Else{
        \texttt{io\_uring.WaitCompletion(frontier)}\;
        \texttt{frontier.SaveResult()}\;
    }
    \texttt{frontier = frontier.next}\;
\end{algorithm}

At the heart of \foreactor is a pre-issuing algorithm that controls I/O speculation on foreaction graphs, designed according to the principles listed in \S\ref{sec:principles-of-speculation}. This algorithm is invoked upon every I/O system call interception when a registered function is active. Pseudocode of the algorithm is given in algorithm~\ref{alg:pre-issuing-algo}.

We would like to point out the critical steps of the algorithm that reflect our design principles. We denote the current intercepted syscall node as the \textit{frontier} node. First of all, this algorithm is invoked upon every intercepted syscall node and stops after a certain depth of peeking beyond this node as line~\ref{alg:line:outer-while} suggests, which corresponds to the speculate-at-run-time principle. Second, peeking is done in code execution order by always following outgoing edges in line~\ref{alg:line:peek-next}. Third, argument values are computed explicitly in line~\ref{alg:line:compute-args}. Fourth, line~\ref{alg:line:track-weak} keeps track of whether there are any weak edges along the path, and line~\ref{alg:line:foreactable-cond} permits a non-pure syscall node being speculatively issued only if there are no weak edges on the path from frontier. Lastly, prepared system calls are submitted to \iouring as a batch of entries in line~\ref{alg:line:batch-submit} to reduce the number of actual user-kernel boundary crossings.

\subsection{Correctness Guarantees}

Since plugin files are compiled directly with the application, they have the visibility of application states and the ability to change them. If the following two rules are obeyed when composing a plugin file, then it is guaranteed that \foreactor will maintain correctness of data during the process of speculation and will not corrupt the application's original semantic.

\paratitle{Keep input variables read-only} All application state variables that appear in the \textit{Input} annotation should remain read-only throughout the plugin file. The only exceptions are destination buffers of \texttt{fstat(at)}, \texttt{(p)read}, and \texttt{getdents} system calls, which for sure would be modified by the original function anyway.

\paratitle{Obey external synchrony} Nightingale et al. have defined the correctness model of \textit{external synchrony} for local file I/O in~\cite{xsyncfs}. Specifically, a system that deploys asynchronous I/O is externally synchronous if an external observer, in this case the user of the application, cannot distinguish the output of the system from its original synchronous counterpart. In our case, if all syscall nodes in a foreaction graph are \textit{pure}, i.e. read-only, this is trivially true. If there are non-pure syscall nodes, then the developer must ensure that the foreaction graph structure matches application code and that all \textit{weak} edges are marked correctly. If so, a non-pure system call instance can be speculatively issued only if it is guaranteed to be invoked, thus not leaving any unwanted side effects that the original synchronous version would not make.


\subsection{Implementation}

We implement \foreactor with 7.5k lines of C++ code, excluding any application-specific plugin files. The length of a plugin file may vary from tens to hundreds of lines, depending on the complexity of its foreaction graph.

\paratitle{Function Call Interception}. \foreactor intercepts chosen application functions through a linker \texttt{--wrap} option trick. Given function \texttt{f}, this option allows us to supply a wrapper function named \texttt{\_\_wrap\_f()} and invoke the original \texttt{f} through symbol \texttt{\_\_real\_f()}. Alternatively, developers could directly inject wrapper code in-place around candidate functions.

\paratitle{System Call Interception} Libforeactor is required to be \texttt{LD\_PRELOAD}ed when running the application, in order to intercept POSIX library calls corresponding to I/O system calls. Currently supported system calls include \texttt{open(at)}, \texttt{close}, \texttt{(p)read}, \texttt{(p)write}, \texttt{fstat(at)}, and \texttt{getdents}.

\paratitle{Asynchronous Backend Engine} By default, \foreactor uses Linux \iouring as the backend asynchronous system call engine. It also supports a built-in user-level thread pool engine with the same semantic as \iouring as an alternative to support older kernel versions and to improve debuggability.

\section{Evaluation}
\label{sec:evaluation}

We evaluate \foreactor with four application use cases: \texttt{du} -- an \texttt{fstat}-heavy utility that calculates storage space usage of a directory, \texttt{cp} -- a \texttt{read}-\texttt{write}-heavy utility that copies files, BPTree -- a standard B+-tree data structure, and LevelDB -- an LSM-tree-based key-value store. We show the effectiveness of \foreactor by answering the following questions:

\begin{itemize}
    \item How well does \foreactor improve the performance of regular I/O request loops? (\S\ref{sec:eval-cli-utilities})
    \item How well does \foreactor apply to more complex application functions, such as LSM-tree operations, with different workload configurations? (\S\ref{sec:eval-bplus-tree} and \S\ref{sec:eval-leveldb-get})
    \item How much is the performance overhead of explicit speculation and how much does it improve device bandwidth utilization? (\S\ref{sec:eval-overhead-util})
\end{itemize}


\paratitle{Experimental Setup} We run all experiments on an Intel x86 server machine with one Toshiba NVMe NAND-flash SSD. The machine is equipped with an Intel Xeon D-1548 dual-socket 8-core CPU running at 2.00GHz (with 2 hyperthreads per physical core) and 64GB of DDR4 ECC main memory. The NVMe SSD device is attached through PCIe v3.0 and has a capacity of 256GB. It offers a bandwidth of 1200MB/s for sequential accesses, 1115MB/s for mixed random 64KB requests at a queue depth (QD) of 16, and 60MB/s for mixed random 4KB requests at a QD of 1. The backing file system is ext4 mounted in default ordered mode.

\subsection{Command-line Utilities}
\label{sec:eval-cli-utilities}

\begin{figure}[t]
    \centering
    \hfill
    \begin{subfigure}[t]{0.94\columnwidth}
        \centering
        \includegraphics[width=0.65\columnwidth]{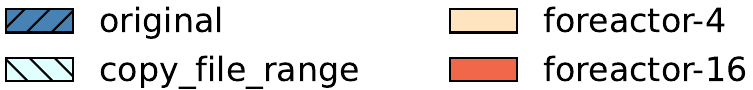}
    \end{subfigure}
    \begin{subfigure}[t]{0.445\columnwidth}
        \includegraphics[width=\columnwidth]{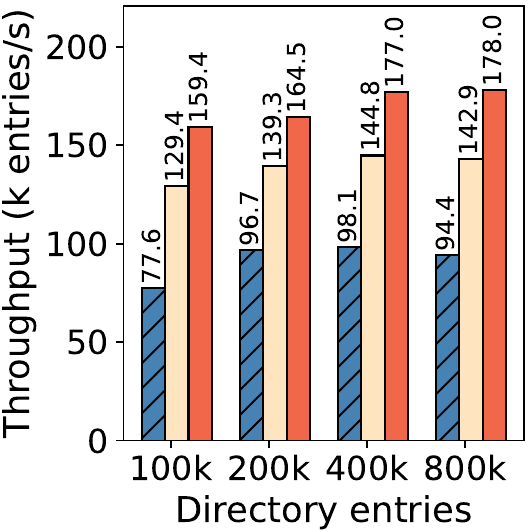}
        \captionsetup{justification=centering}
        \caption{\texttt{du} Stat Loop}
        \label{fig:eval-cli-du}
    \end{subfigure} \hfill
    \begin{subfigure}[t]{0.535\columnwidth}
        \includegraphics[width=\columnwidth]{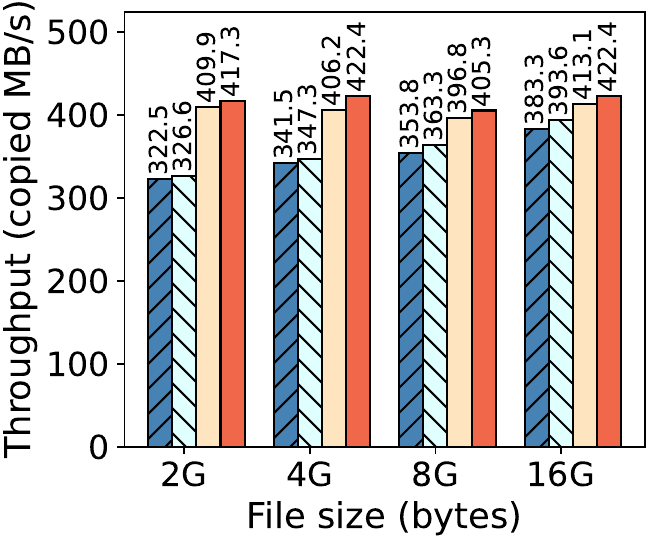}
        \captionsetup{justification=centering}
        \caption{\texttt{cp} Copy Loop}
        \label{fig:eval-cli-cp}
    \end{subfigure}
    \complexcap{Throughput of \texttt{du} and \texttt{cp} Workloads}{}
    \label{fig:eval-cli-utilities}
    \vspace{-5pt}
\end{figure}

\begin{figure}[!t]
    \centering
    \hfill
    \begin{subfigure}[t]{0.94\columnwidth}
        \centering
        \includegraphics[width=0.8\columnwidth]{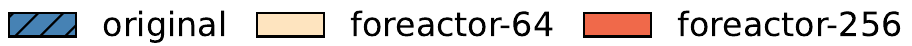}
    \end{subfigure}
    \begin{subfigure}[t]{0.49\columnwidth}
        \includegraphics[width=\columnwidth]{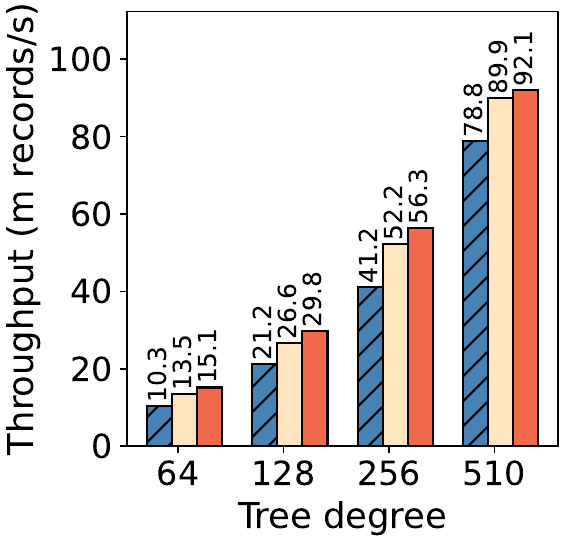}
        \captionsetup{justification=centering}
        \caption{Scan}
        \label{fig:eval-bpt-scan}
    \end{subfigure} \hfill
    \begin{subfigure}[t]{0.49\columnwidth}
        \includegraphics[width=\columnwidth]{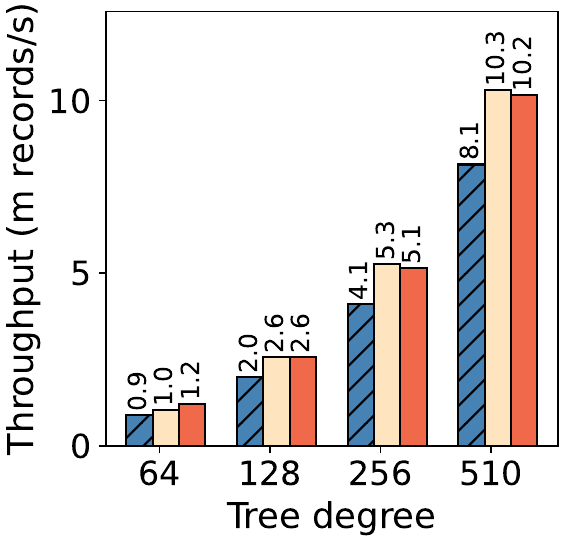}
        \captionsetup{justification=centering}
        \caption{Bulk-loading}
        \label{fig:eval-bpt-load}
    \end{subfigure}
    \complexcap{Throughput of BPTree \texttt{Scan} and \texttt{Load}}{}
    \label{fig:eval-bplus-tree}
    \vspace{2pt}
\end{figure}

\begin{table}[!t]
    \centering
    \begin{tabular}{c|c|c}
        \hline
        \textbf{Throughput} (m rec.s/s) & \textbf{\texttt{io\_uring}} & \textbf{User threads} \\ \hline
        Scan  (degree = 510) & 92.1 & 83.1 \\ \hline
        Load  (degree = 510) & 10.2 & 9.5 \\ \hline
    \end{tabular}
    \complexcap{BPTree Performance on Different Backends}{}
    \label{tab:eval-bpt-backends}
    \vspace{-12pt}
\end{table}

To test the effectiveness of explicit speculation on regular I/O request loops, we consider two command-line utilities, \texttt{du} and \texttt{cp}, that are part of the GNU \texttt{coreutils} suite. \texttt{du} is used for calculating disk space usage of a directory by calling \texttt{fstat}s on its entries to gather their file sizes. \texttt{cp} is used for copying files between different locations in the file system directory tree. On recent Linux kernels, \texttt{cp} also adopted the new \texttt{copy\_file\_range} interface to reduce the number of system calls and to avoid user buffer copies if the backing file system supports it. We compare against this mode as well.


\begin{figure*}[t]
    \centering
    \begin{subfigure}[t]{\textwidth}
        \centering
        \includegraphics[width=0.27\textwidth]{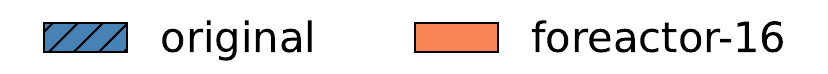}
    \end{subfigure}
    \begin{subfigure}[t]{0.33\textwidth}
        \includegraphics[width=\columnwidth]{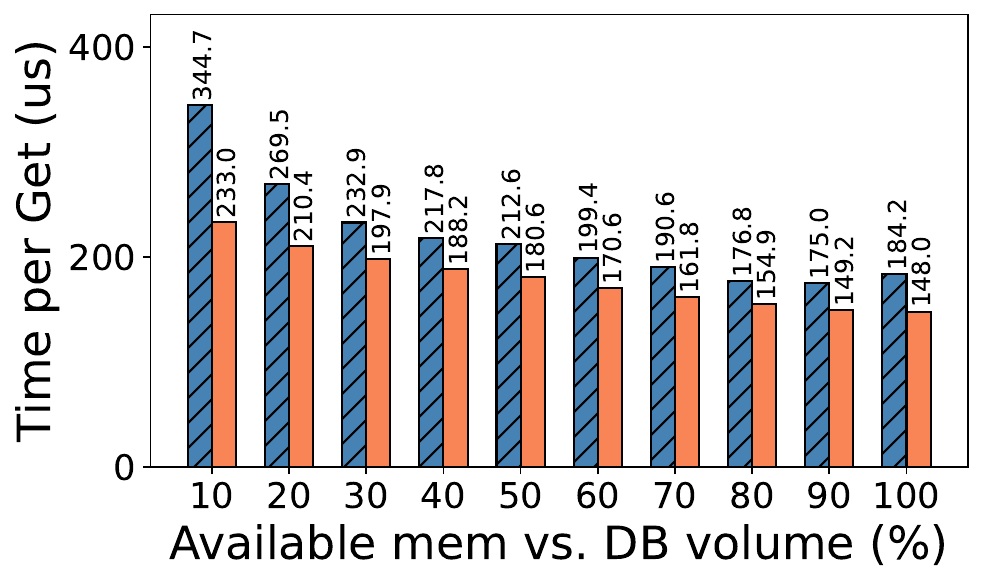}
        \captionsetup{justification=centering}
        \caption{Avg. Time vs. Memory Ratio\\Record Size = 1K}
        \label{fig:eval-ldb-mem-ratio}
    \end{subfigure} \hfill
    \begin{subfigure}[t]{0.33\textwidth}
        \includegraphics[width=\columnwidth]{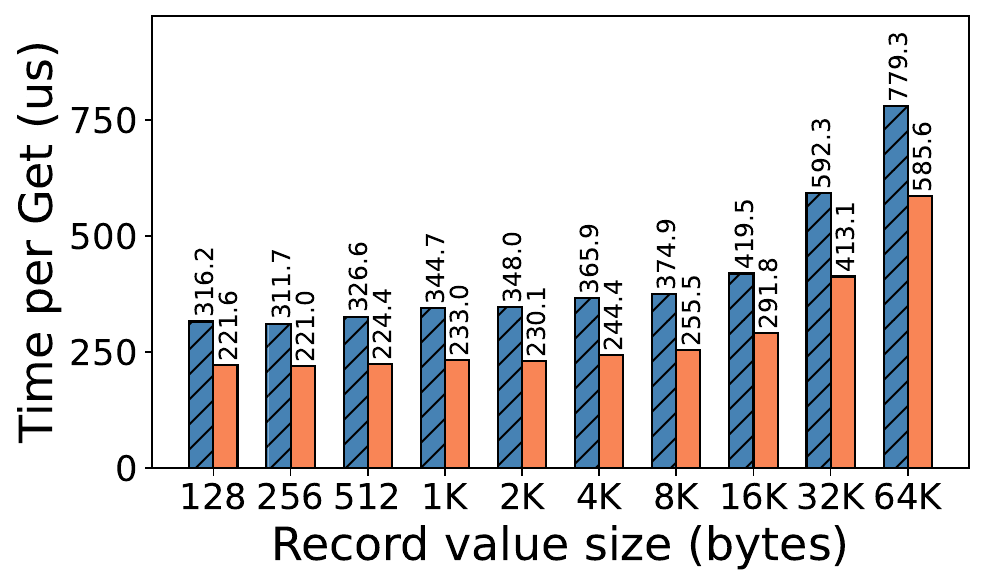}
        \captionsetup{justification=centering}
        \caption{Avg. Time vs. Record Size\\Memory Ratio = 10\%}
        \label{fig:eval-ldb-req-size}
    \end{subfigure} \hfill
    \begin{subfigure}[t]{0.33\textwidth}
        \includegraphics[width=\columnwidth]{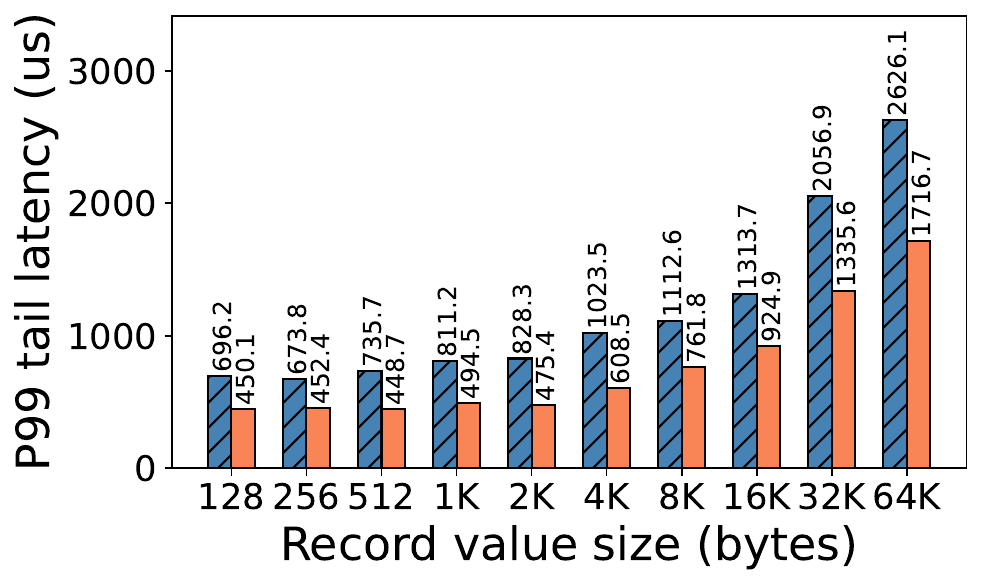}
        \captionsetup{justification=centering}
        \caption{P99 Tail Latency vs. Record Size\\Memory Ratio = 10\%}
        \label{fig:eval-ldb-tail-lat}
    \end{subfigure}
    \complexcap{LevelDB Get Operation Performance}{\subref{fig:eval-ldb-mem-ratio} and \subref{fig:eval-ldb-req-size}: avg. time to complete a \texttt{Get} operation in YCSB-C workload with different page cache memory capacities and record sizes; \subref{fig:eval-ldb-tail-lat}: 99\% tail latency versus record sizes.}
    \label{fig:eval-ldb-avg-time}
    \vspace*{-5pt}
\end{figure*}

Figure~\ref{fig:eval-cli-du} shows the average completion time of a single invocation of \texttt{du} on directories containing different numbers of files. Within each group, the left bar corresponds to original \texttt{du}, while the right two bars correspond to \texttt{du} with \foreactor parallelizing its \texttt{fstat} loop at different pre-issuing depths, 4 and 16. Each bar is averaged over 20 runs. The kernel VFS inode cache is emptied before each run, since \texttt{du} is typically run on cold directories. We can see that maintaining multiple metadata reads in parallel reduces completion time by up to 50\%, thanks to concurrent requests multiplexing device bandwidth. We also observe better performance with larger pre-issuing depth, though the trend flattens out due to higher overhead running the pre-issuing algorithm.

Figure~\ref{fig:eval-cli-cp} shows the average completion time of a single invocation of \texttt{cp} on files of different sizes. Each bar is averaged over 20 runs and the kernel page cache is emptied before each run. \foreactor outperforms original \texttt{cp} and its \texttt{copy\_file\_range} mode, though there is only marginal benefit due to two reasons. First, each iteration of the copy loop issues a read request of 128KB size and a write of the same size, which is already a large enough scale to utilize device bandwidth. Second, we ran \texttt{cp} with file source and destination on the same backend device, introducing read-write interference. In a data backup workload where source and destination are on different storage devices, this effect will be minimized.


\subsection{B+-tree Scan and Bulk-loading}
\label{sec:eval-bplus-tree}

We implement BPTree, a standard B+-tree data structure that supports \texttt{Scan} and \texttt{Load} operations. We then apply \foreactor to parallelize the leaf page I/O loops in these two operations as described in \S\ref{sec:app-study-bplus-tree}. We use 8KB as the page size and consider 64-bit integral keys and values. A BPTree instance has a controllable \textit{degree} (i.e., the number of keys a node can hold before triggering a split) configured at initialization, with a maximum degree of 510 allowed in our settings. A Higher degree leads to a wider tree and in turn fewer I/O requests.

We run \texttt{Load} to build a BPTree instance from a pre-generated key-value record stream followed by 10 range \texttt{Scan}s, and repeat this for different degree configurations. All keys are chosen uniformly random. Figure~\ref{fig:eval-bpt-scan} and \ref{fig:eval-bpt-load} shows the throughput of \texttt{Scan} and \texttt{Load}, respectively, in million records per second with different degree configurations. \foreactor is able to speculatively issue leaf page I/O requests, bringing up to 37\% of throughput improvement. We can also see that a B+-tree with higher degree offers better bulk operation performance, because there will be a significantly fewer number of nodes. The cost is slower binary search for a key within a node, which is negligible for bulk operations but may be significant for point queries. Note that we choose a pre-issuing depth of up to 256 versus 16 for \texttt{cp}, because the request sizes are 8KB pages instead of 128KB buffers.


Table~\ref{tab:eval-bpt-backends} compares BPTree performance on two different asynchronous I/O backends: the default \texttt{io\_uring} and the user-level thread pool. \texttt{io\_uring} delivers better throughput due to a reduction in the number of actual system calls and the efficiency of the kernel \texttt{io\_}workqueue.

\subsection{LevelDB Get with YCSB}
\label{sec:eval-leveldb-get}

\begin{figure*}[t]
    \centering
    \captionsetup{}
    \begin{minipage}{0.31\textwidth}
        \centering
        \includegraphics[width=\linewidth]{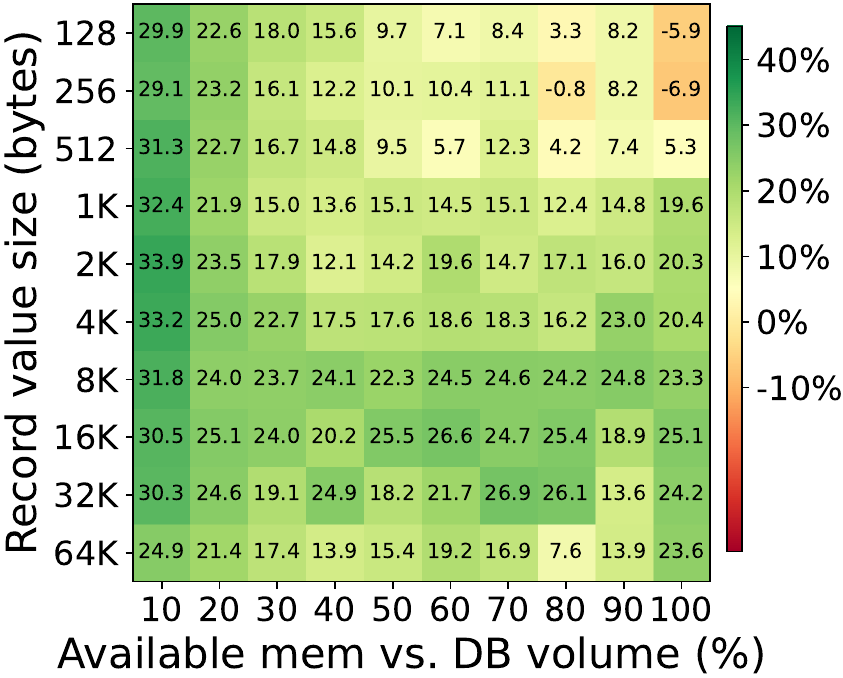}
        \vspace*{-15pt}
        \captionsetup{justification=raggedright}
        \complexcapoffig{LevelDB Get Relative Improvement}{Greener is better.}
        \label{fig:eval-ldb-heat-map}
    \end{minipage}%
    \begin{minipage}{0.69\textwidth}
        \centering
        \begin{subfigure}[t]{0.33\linewidth}
            \includegraphics[width=\columnwidth]{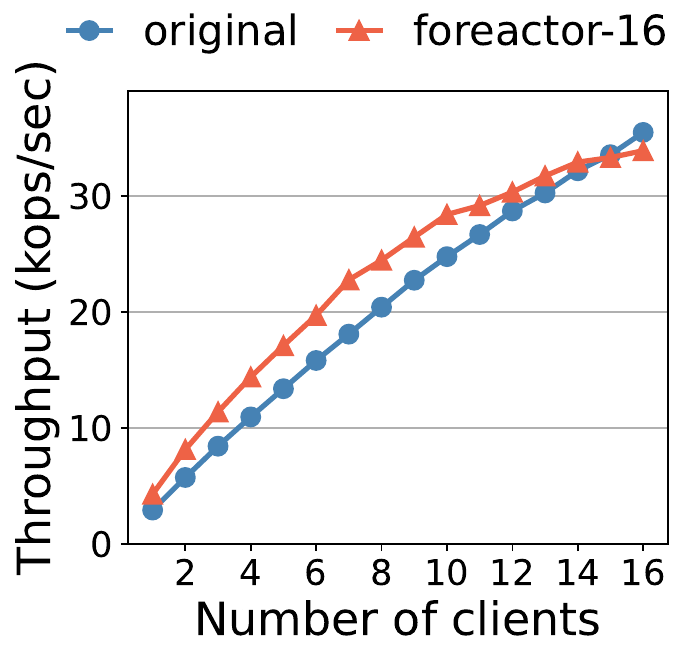}
            \captionsetup{justification=centering}
            \caption{Multiple Clients}
            \label{fig:eval-ldb-multithread}
        \end{subfigure} \hfill
        \begin{subfigure}[t]{0.325\linewidth}
            \includegraphics[width=\columnwidth]{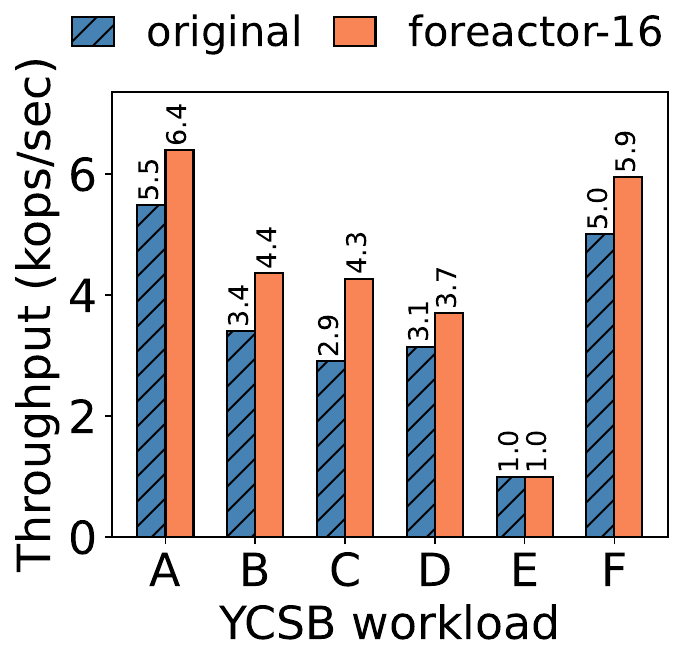}
            \captionsetup{justification=centering}
            \caption{YCSB Workloads}
            \label{fig:eval-ldb-with-writes}
        \end{subfigure} \hfill
        \begin{subfigure}[t]{0.33\linewidth}
            \includegraphics[width=\columnwidth]{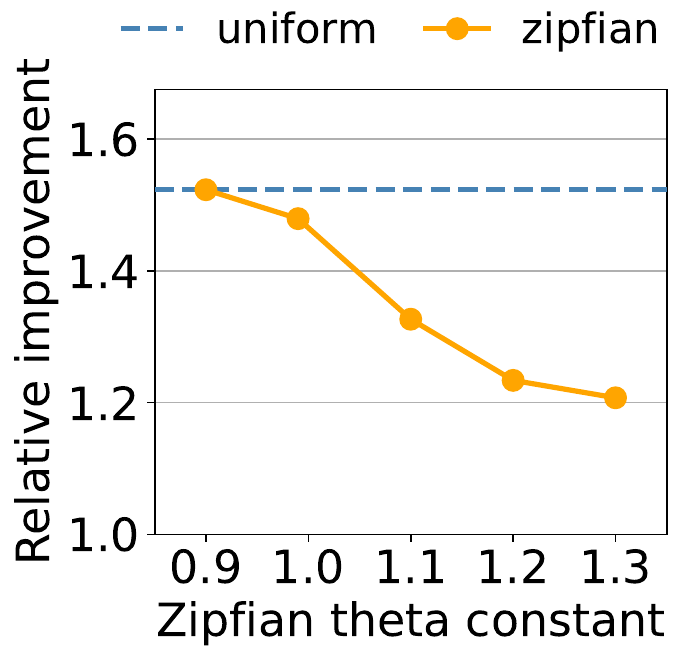}
            \captionsetup{justification=centering}
            \caption{Workload Skewness}
            \label{fig:eval-ldb-zipf-consts}
        \end{subfigure}%
        \addtocounter{figure}{-1}  
        \complexcapoffig{LevelDB Performance Sensitivity Analysis}{Throughput under the impact of multiple clients, non-\texttt{Get} operations, and different key skewness.}
        \label{fig:eval-ldb-sensitivity}
    \end{minipage}
\end{figure*}

To apply \foreactor to a more complex use case, we choose LevelDB, an LSM-tree-based key-value store. In particular, we use \foreactor to parallelize the \texttt{pread}s in its \texttt{Get} request code path described in \S\ref{sec:app-study-lsm-tree}.

We run YCSB-C workload with default Zipfian distribution on LevelDB with different page cache memory capacities (set through \texttt{cgroup}) and record value sizes. In each experiment, we run a workload trace that reads out 200MB of data from a database image of approximately 1GB volume size. We choose a pre-issuing depth of 16 as it is the upper bound of the number of \texttt{pread}s in our settings. Results are presented in Figure~\ref{fig:eval-ldb-avg-time}. Each bar is averaged over 5 runs. Figure~\ref{fig:eval-ldb-mem-ratio} shows that \foreactor is able to reduce average \texttt{Get} latency by speculatively issuing future \texttt{pread}s. The improvement gets better with smaller page cache memory ratio, because an increasing number of \texttt{pread}s will miss the page cache and land on the SSD device. Figure~\ref{fig:eval-ldb-req-size} shows that for small record sizes below 4KB, \foreactor brings steady improvement because all \texttt{pread}s will actually read out 4KB data blocks. For larger value sizes, the improvement tends to get better due to each \texttt{pread} getting larger, outweighing the computation overhead. Figure~\ref{fig:eval-ldb-tail-lat} shows that explicit speculation is especially helpful for improving tail latency of LevelDB \texttt{Get}s, because the slowest \texttt{Get}s are ones with the longest search chains and that trigger the most page cache misses.

Figure~\ref{fig:eval-ldb-heat-map} presents a summary of relative \texttt{Get} operation performance improvement across all memory ratios and record sizes. There is a clear trend that \foreactor helps the most with a small memory ratio and a moderate record size. The top-right corner shows that it is not wise to apply \foreactor when memory capacity is abundant and request size is small, because then most \texttt{pread}s will be small memory accesses hitting the page cache.

\paratitle{Sensitivity Analysis} We run more complicated workloads to study the effectiveness of \foreactor when there are multiple LevelDB client threads, when the workload is a mix of read and write operations, and with different workload skewness. Figure~\ref{fig:eval-ldb-multithread} shows that \foreactor scales well when the number of clients are small, because each LevelDB client thread performs explicit speculation independently and uses its own \iouring queue pair. When too many clients co-locate on the same machine, device bandwidth is already sufficiently utilized and \foreactor starts to behave poorly due to bandwidth contention and excessive computation overhead. Figure~\ref{fig:eval-ldb-with-writes} shows single-client throughput of different YCSB workloads while only \texttt{Get} operations are being accelerated by \foreactor. Improvement is proportional to the percentage of \texttt{Get} requests in each workload. Figure~\ref{fig:eval-ldb-zipf-consts} studies the effect of search key distribution skewness other than default Zipfian constant 0.99. Smaller Zipfian constant corresponds to less skew workload and higher chance of page cache misses, resulting in better relative improvement.

\subsection{Overhead and Utilization}
\label{sec:eval-overhead-util}

\begin{figure}[t]
    \centering
    \includegraphics[width=\columnwidth]{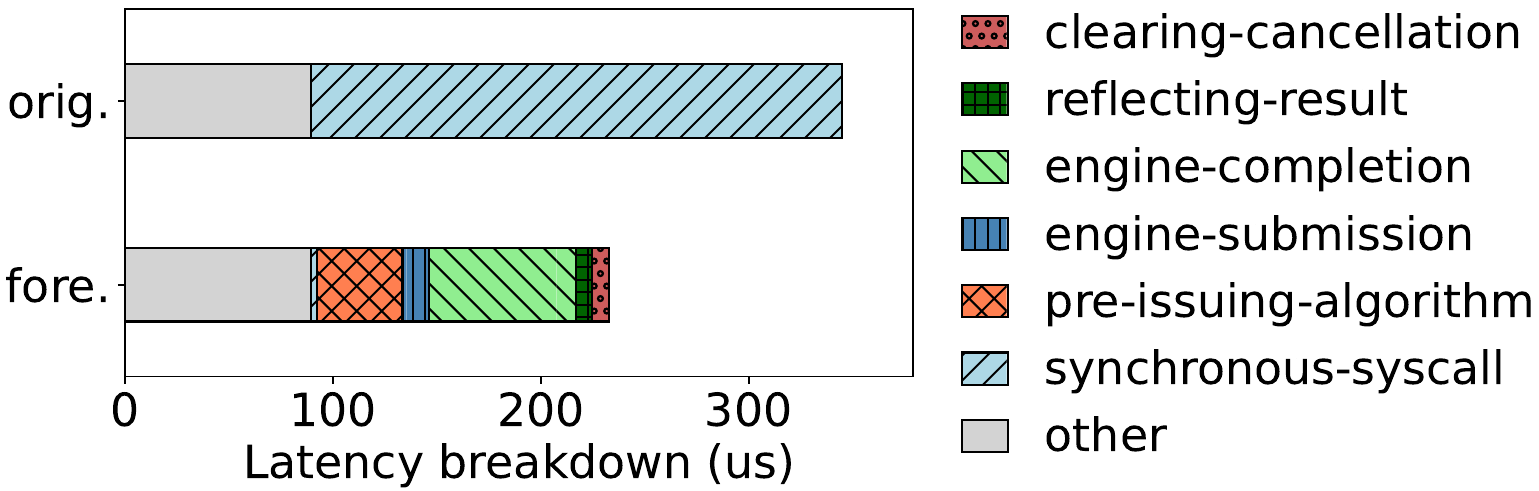}
    \complexcap{Latency Breakdown of LevelDB Get Case}{}
    \label{fig:eval-ldb-breakdown}
    \vspace{-8pt}
\end{figure}

\foreactor introduces three pieces of performance overhead in order to enable explicit speculation: \circlew{1} executing the pre-issuing algorithm upon intercepted system calls, \circlew{2} copying read system call results back to user buffers, and \circlew{3} cancelling on-the-fly system calls upon early exits. Figure~\ref{fig:eval-ldb-breakdown} quantifies the latency factors of a typical LevelDB Get operation at 1KB record size and 10\% memory ratio, with and without \foreactor. Gray bars correspond to irrelevant application logic not touched by \foreactor; light-blue bars correspond to synchronous I/O system calls, in this case \texttt{pread}s; blue and light-green bars represent the time spent in submitting \texttt{pread}s to \iouring and waiting on their completions, which is where \foreactor saves time compared to original synchronous system calls. The rest three bars correspond to overhead factors, with pre-issuing algorithm being the largest factor. In general, \foreactor brings improvement as long as the benefit of I/O concurrency outweighs the overhead of peeking.

A small amount of extra CPU power and memory space is required for adding I/O concurrency. Quantitatively, for LevelDB Get with 1KB record size, applying \foreactor to a single client thread increases CPU usage on our 8-core machine from 11.95\% to 14.75\% and allocates roughly 1.06MB of extra memory for read buffers. As a result, SSD device bandwidth utilization boosts from 130.60MB/s to 197.43MB/s per client thread. Other workloads report similar numbers.

\section{Discussion}
\label{sec:discussion}

We provide additional discussion on some practical aspects of applying explicit speculation with Foreactor, specifically, on the effects of application-level caching, OS page cache configurations, and possible automation in generating foreaction graphs. We also discuss the limitations of explicit speculation.

\paratitle{Effects of Application-level Caching} For applications that manage a built-in cache atop I/O system calls, its foreaction graphs might be less straightforward to draw, since \texttt{read} and \texttt{write} requests might hit the application-level cache and hence do not trigger corresponding system calls. In this case, one solution is to turn off the application-level cache and instead use the OS-level page cache. A possible extension to the foreaction graph abstraction is to generalize the definition of syscall nodes and allow them to also represent requests to the application-level cache (which may or may not translate into actual system call instances). This extension may require modifications to application source code to let \foreactor take control at cache query boundaries in addition to intercepting system calls, and is currently not implemented in \foreactor.

\paratitle{Effects of OS Page Cache} Since \iouring is mostly oblivious to the Linux kernel page cache, non-\texttt{O\_DIRECT} I/O system calls are naturally supported. Depending on the workload and available memory capacity, it might be desirable to turn off page cache read-ahead by \texttt{fadvise(FADV\_RANDOM)} and/or \texttt{fadvise(FADV\_DONTNEED)} so as to reduce the chance of page cache contention by concurrent I/O requests. Another pitfall is that the Linux page cache locks all accesses to a file exclusively during buffered writes (through the \texttt{rwsem} semaphore), which discourages concurrent write requests to the same file even when all writes are to non-overlapping address ranges. For such workloads, the \texttt{RWF\_UNCACHED} flag might be useful to avoid page cache thrashing~\cite{rwf-uncached}.

\paratitle{Obtaining Foreaction Graphs} So far, we have assumed that foreaction graphs are derived manually by application developers with sufficient understanding of application code logic. It is possible to automate (part of) this process through static code analysis by compilers. For example, the graph structure could be derived from the control-flow graph (CFG) of the function, which is an intermediate representation during the compilation procedure, by extracting out all occurrences of system calls. Annotations could be generated by semantic analysis on code, through techniques such as symbolic execution. We leave automatic generation of foreaction graphs as future work to explore.

\paratitle{Limitations} The explicit speculation approach currently has two major limitations. Firstly, it does not help in cases where I/O requests form a strict dependency chain, i.e., each request depends on the result of its previous request. A typical example would be the pointing-chasing procedure of a B+-tree search. In this case, none of the later \texttt{read}s can be pre-issued concurrently, unless we could make precise guesses on the addresses of lower-level blocks. We would like to note that explicit speculation can be freely combined with other I/O acceleration techniques. For example, XRP -- I/O request aggregation with eBPF embedding~\cite{xrp} -- can be applied to such cases. The second limitation is that the foreaction graph abstraction is not expressive enough to represent complex recursive functions, since it is based on loops and does not track the function call stack. The scope of I/O speculation is thus limited to non-recursive functions, which we found expressive enough to cover many interesting cases.

\section{Related Work}
\label{sec:related-work}

There has been a rich set of work on measurement of storage device characteristics, software I/O acceleration techniques, and application system call analysis.

\paratitle{Device Parallelism Measurement} He et al. presented a performance study of NAND flash-based SSDs~\cite{unwritten-contract-ssd} and showed that maintaining a large aggregate request scale is necessary to fully utilize device bandwidth due to the fact of internal stripping across smaller flash media units. Wu et al. presented a similar study on Intel Optane SSDs~\cite{unwritten-contract-optane}, a recent type of SSDs based on the 3D-XPoint persistent memory media, and also reported the effect of device internal parallelism, though to a smaller degree compared to NAND flash SSDs. There are also studies applying RAID to SSDs to further improve parallelism and reliability~\cite{raid-ssd, diff-raid}. Overall, the effect of storage I/O parallelism is common on modern storage hardware.

\paratitle{I/O Speculation} As described in detail in \S\ref{sec:io-speculation}, previous I/O speculation techniques have taken either the speculative execution approach~\cite{spechint-kernel, speculator, speck, xsyncfs, specrpc} for compute- or write-heavy workloads, or the speculative prefetching approach~\cite{spechint, specprefetch, informed-prefetch, sig-based-prefetch, compiler-inserted-prefetch} for read-only cache workloads. This paper proposes explicit speculation, a more deterministic and formal approach, to support controlled I/O speculation with an emphasis on exploiting storage I/O parallelism.

\paratitle{Asynchronous I/O} Linux \iouring is a new asynchronous I/O interface supported in kernel version 5.1 or newer~\cite{io-uring}, with which users can submit original I/O system calls in batches and poll on their completions using the minimum number of actual user-kernel boundary crossings. It exposes a dual ring-queue programming interface and implements an efficient in-kernel \texttt{io\_}workqueue thread pool. Significant engineering effort is still required to enable asynchronous I/O in mature applications, for example, in scientific computing applications~\cite{async-io-vol, async-io-hpc} and databases~\cite{async-iterator-model}.

\paratitle{Other I/O Acceleration Techniques} A wide range of I/O acceleration techniques have been proposed to reduce storage stack software overhead in the era of ultra-fast storage devices. Examples include caching and memory mapping~\cite{orthus, fastmap}, batching and aggregation~\cite{cassyopia, multicall, flexsc, xrp, zio}, kernel-bypassing~\cite{strata, arrakis, aerie, ufs, demikernel}, disaggregation~\cite{devfs, crossfs, fusionfs, legoos}, and device-specific systems such as persistent memory file systems~\cite{dax, nova, splitfs, zofs, odinfs}. Explicit speculation proposed in this paper is an orthogonal aspect to these techniques.

\paratitle{Application System Call Analysis} Previous works have used system call analysis to study the correctness of applications and storage system semantics. For example, ALICE~\cite{alice} is a tool that analyzes application crash consistency behavior on different file system persistence models. Frost et al. proposed a file system abstraction, patch, that explicitly encodes write-before dependencies in their consistency mechanisms and allows lower layers of the storage stack to be file-system-agnostic~\cite{generalized-fs-dependencies}. \foreactor expands the scope of system call analysis to build foreaction graphs and improve performance.

\section{Conclusion}
\label{sec:conclusion}

In this paper, we introduce explicit speculation, a highly deterministic variant of I/O speculation technique. We propose the foreaction graph abstraction that describes application function I/O patterns and captures any necessary computation required to produce their argument values. We implement \foreactor, a framework that allows application developers to materialize foreaction graphs and exploit storage I/O parallelism transparently. Experimental results show that \foreactor is able to improve device bandwidth utilization and boost performance across a range of I/O-intensive applications.


\bibliographystyle{plain}
\bibliography{references}


\end{document}